%% file: paper.tex
\documentclass[conference]{IEEEtran}
  
\usepackage{stmaryrd}
\usepackage{amsfonts}
\usepackage{flushend}
\usepackage{subcaption}
\usepackage{flushend}
\usepackage{epstopdf}
\usepackage{booktabs}
\usepackage{multirow}

\usepackage{graphicx,times,amsmath}

\IEEEoverridecommandlockouts  

\usepackage{fancyheadings,color}
\title{Comparing LBP, HOG and Deep Features\\ for Classification of Histopathology Images}

\author{%
 \vspace{0.1in} Taha J. Alhindi$^{1,2,3}$,  Shivam Kalra$^{*1,3}$, Ka Hin 
  Ng$^3$, Anika Afrin$^4$, Hamid R. Tizhoosh$^{1}$\\ 
   $^1$ \textit{Kimia Lab, University of Waterloo, Canada}\\ 
  $^2$ \textit{Dept. of Industrial Eng., King Abdulaziz Univ., Jeddah, Saudi Arabia}\\
  $^3$ \textit{Systems Design Engineering, University of Waterloo, Canada}\\
  $^4$ \textit{Electrical and Computer Engineering, University of Waterloo, Canada}\\
  {\textit{\{shivam.kalra, talhindi, kh7ng, aafrin, tizhoosh\}@uwaterloo.ca}}\\ 
  \thanks{$^*$Shivam Kalra is a corresponding author for the research work.}
}

\begin{document}
\maketitle

\begin{abstract}
Medical image analysis has become a topic under the spotlight in recent
years. There is a significant progress in medical image research concerning the
usage of machine learning. However, there are still numerous questions and
problems awaiting answers and solutions, respectively. In the present study,
comparison of three classification models is conducted using features extracted
using local binary patterns, the histogram of gradients, and a pre-trained deep network. Three common image classification methods, including support
vector machines, decision trees, and artificial neural networks are used to
classify feature vectors obtained by different feature extractors. We use KIMIA
Path960, a publicly available dataset of $960$ histopathology images extracted
from $20$ different tissue scans to test the accuracy of classification and feature
extractions models used in the study, specifically for the histopathology
images. SVM achieves the highest accuracy of $90.52\%$ using local binary
patterns as features which surpasses the accuracy obtained by deep features, namely $81.14\%$.
\end{abstract}

\section{Introduction}

\PARstart{I}n recent years, machine learning has become a popular topic and its
applications are increasing day by day with respect to image-based diagnosis,
disease prediction and risk
assessment~\cite{deBruijneMachinelearningapproaches2016}. Machine learning is
considered a sub-area of artificial intelligence, a set of methods which learn
from past data to generalize to new data, can handle noisy input and complex
data environments, use prior knowledge, and can form new
concepts~\cite{AIItsNature2016}.

After recent success of machine learning, and specially deep learning, in various
application fields, some approaches are providing solutions with good accuracy
for medical imaging making it a great opportunity for future applications in the
healthcare sector. Major experimentation attempts for computer-aided diagnosis
had begun in the mid-1980s when the primary focus was on the techniques for
detecting lesions on chest radiographs and
mammograms~\cite{GigerAnniversarypaperHistory2008}. In recent years,
machine-learning approaches are being used successfully in the area of
image-based disease detection and forecasting. Oliver et al. proposed an
approach for reducing false positives for recognition of mammography images
using local binary pattern (LBP) in
2007~\cite{OliverFalsePositiveReduction2007}. LBP was used for extracting the
descriptors and detected masses were classified into either malignant or benign
with support vector machines (SVM). The result of the experiment showed that the
LBP features were successful in not only in terms of false positive reduction
but also it was efficient compared to other methods for diverse mass areas,
which is an acute feature of the systems for mass detection. An appraisal of
bag-of-features approach for classifying histopathology images had been proposed
by Caicedo et al. in 2009~\cite{CaicedoHistopathologyImageClassification2009}.
The key advantage of their proposed framework is that it is focusing to the
contents of the image group particularly. This property is achieved by an
automated codebook construction and the visual feature descriptors.

An image analysis methodology using SVM classifier to differentiate low and high
grades of breast cancer automatically has been proposed by Doyle et al. in
2008~\cite{DoyleAutomatedgradingbreast2008}. The dataset contained 48 breast
biopsy tissue images having over 3400 image features. After the feature
dimensionality reduction and SVM classification, the system achieved an accuracy
of $95.8\%$ in differentiating cancerous from non-cancerous cases where Gabor
filter features had been used. Distinguishing high-level from low-level cancer
was done with architectural features and an accuracy of $93.3\%$ was achieved.
Moreover, they used spectral clustering to visualize the hidden manifold form
which consists various grades of cancer, and it showed a steady shift from
low-grade to high-grade breast cancer.

Kumar et al. conducted a recent study which compares deep features,
bag-of-visual-words (BoVW) and LBP for the classification of histopathology
image dataset, KIMIA Path960. The classification accuracy obtained in the study
using LBP and deep features were $90.62\%$ and $94.72\%$, respectively, whereas
BoVW achieved the highest accuracy of $96.50\%$~\cite{KumarComparativeStudyCNN2017}.

In the present study, comparison of three classification models is conducted.
The dataset used is KIMIA Path960~\cite{KumarComparativeStudyCNN2017}. The
models use one of the following feature extractors: Local binary pattern (LBP),
histogram of gradients (HOG) and deep features from VGG 19, a pre-trained deep
network. The feature vectors are then provided as an input to train several
classifiers, for this paper, SVM, decision trees (DTs), and Artificial Neural
Networks (ANNs, which use a shallow multi-layer-perceptron) have been selected.
The study will empirically justify applications of various feature extraction
models and classification methods that are commonly available for the
histopathology image analysis.

The study is divided into 5 sections, where feature extractors and
classification algorithms are discussed in Section~\ref{sec:background}. The
dataset is introduced in Section~\ref{sec:dataset}. Section~\ref{sec:results}
explains the methodology used in the study. Experiments and results are
discussed in Section~\ref{sec:results} later. Conclusions are drawn in
Section~\ref{sec:conclusion}.

\section{Background}\label{sec:background}
\subsection{Feature Extractors}
\textbf{Local Binary Pattern (LBP):} LBP was introduced in 1994~\cite{OjalaPerformanceevaluationtexture1994}. Yet,
the texture spectrum model of LBP was proposed even earlier in
1990~\cite{HeTextureUnitTexture1990}. LBP feature extractor has been applied in
many areas after Ojala and Pietikainen's research in multi-resolution approach
in 2002~\cite{OjalaMultiresolutiongrayscalerotation2002}, including text
identification~\cite{JungLocalBinaryPatternBased2010a} and face
recognition~\cite{CaicedoHistopathologyImageClassification2009}. Wang has
combined LBP with Histogram of Oriented Gradients (HOG) descriptor to improve
detection performance in~\cite{WangHOGLBPhumandetector2009}.

\begin{figure*}[htb]
  \centering
  \vspace{0.1in}
  \includegraphics[width=0.55\textwidth]{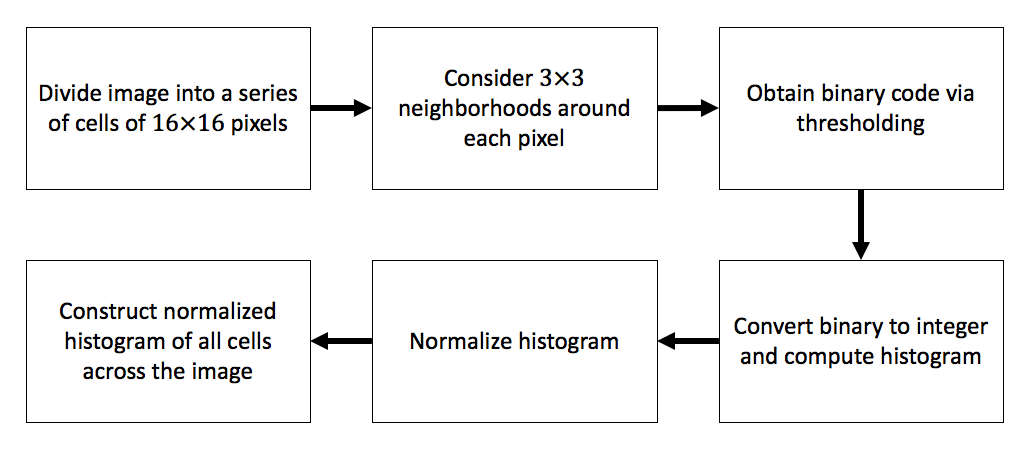}
  \caption{Flowchart summarizing the main steps of LBP feature extractor.\label{fig:lbpflow}}
\end{figure*}

Classical LBP feature extractor should follow specific actions; these actions
are summarized in Fig.~\ref{fig:lbpflow}~\cite{KumarComparativeStudyCNN2017}.

There are series of attempts to improve the LBP method. One recent version finds
the intensity values of points in a circular neighborhood by considering the
values which have small circular neighborhoods around the central
pixel~\cite{BrahnamLocalBinaryPatterns2014}.

\begin{figure*}[ht]
  \centering
  \includegraphics[width=0.55\textwidth]{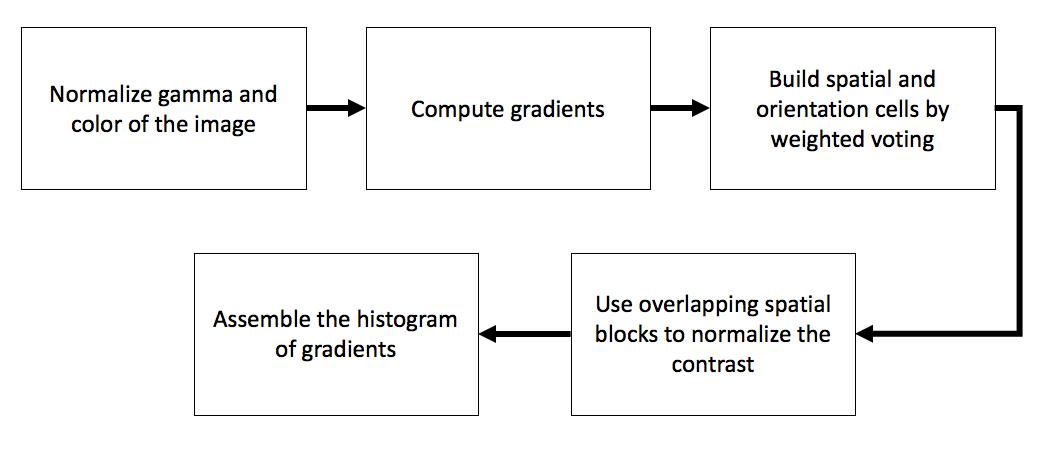}
  \caption{Flowchart summarizing the process of computing HOG.\label{fig:hogflow}}
\end{figure*}

\textbf{Histogram of Oriented Gradients (HOG):} HOG captures features by counting the occurrence of gradient orientation.
Traditional HOG divides the image into different cells and computes a histogram
of gradient orientations over them~\cite{BrahnamLocalBinaryPatterns2014}. HOG is
being applied extensively in object recognition areas as facial
recognition\cite{DenizFacerecognitionusing2011}. The process for computing HOG
is explained in
Fig.~\ref{fig:hogflow}~\cite{DalalHistogramsorientedgradients2005}.

\textbf{VGG:A Pre-Trained Deep Net:} VGG 16 and 19 are deep convolutional networks (ConvNet) architecture first
proposed by K. Simonyan and A. Zisserman from Visual Geometry Group of
University of Oxford in 2014~\cite{SimonyanVeryDeepConvolutional2014}. In the
paper, an evaluation of very-deep networks for large-scale image classification
was carried out: the generic architecture of the network contained a very small
convolution filter with small receptive field of $3\times 3$ and the
convolutional pace was fixed to 1 pixel, while five max-pooling layers carried
out the spatial pooling, over a $2\times 2$ pixel window, with a convolution
step of 2~\cite{SimonyanVeryDeepConvolutional2014}. There were three
fully-connected layers of same configuration: the first and second layers had
4096 channels each, whereas, the third layer contained 1000 channels and
performed ILSVRC-2012 dataset
classification~\cite{SimonyanVeryDeepConvolutional2014}. The ConvNet
configurations analyzed in this study followed the same architecture which only
differ in depth: from network A to E where the depths differ from weight layer
11 to 19~\cite{SimonyanVeryDeepConvolutional2014}. The results of classification
experiments in single-scale, multi-scale and multi-crop evaluation shown that a
major improvement could be achieved in this proposed network if the depth was
pushed to 16 and to 19 weight layers, which were VGG16 and
VGG19~\cite{SimonyanVeryDeepConvolutional2014}.

In recent years, deep learning has been widely used in medical sector in terms
of automated cancer and lesion detection. In 2015, Ertosun and Rubin developed a
search system basing on deep learning to automatically search and localize
masses, where the system contained a classification and a localization engine as
well~\cite{ErtosunProbabilisticvisualsearch2015}. There are also works applying
deep convolution neural network classifiers to mammograms such as
AlexNet~\cite{KrizhevskyImagenetclassificationdeep2012}, VGG
net~\cite{SimonyanVeryDeepConvolutional2014} and
GoogLeNet~\cite{SzegedyGoingDeeperConvolutions2014}.

Another study on finding breast cancer from lymph node biopsy images has been
carried out by Wang, Dayong, et al.~\cite{WangDeepLearningIdentifying2016} in
2016 using deep learning classification models, including GoogLeNet, VGG-16,
AlexNet and FaceNet. Combining estimation from deep learning systems with the
diagnosis of the human pathologist, the pathologist's area under the receiver
operating curve (AUC) was increased to $0.995$, representing nearly $85\%$
reduction in human error rate~\cite{WangDeepLearningIdentifying2016}, which
means that integration of deep learning approaches with the workflow of
pathologists could increase the accuracy in cancer diagnosis.

\textbf{Radon Features:}  Although we do not experiment with them in this study, we have to mention several approaches that have been proposed to use Radon transform for feature extraction \cite{babaie2017local,Tizhoosh2018}. These methods use projections in local neighbourhood to assemble a feature vector or a histogram. The most recent work reports retrieval accuracies using encoded local projections, short ELP histogram, that surpass deep features for histopathology images \cite{Tizhoosh2018}. 

\subsection{Classifiers}
\textbf{Support Vector Machines (SVM):} SVM is a classifier designed for binary problems with extension to multi-class
problems~\cite{BeggComputationalIntelligenceBiomedical2007}. Cortes and Vapnik firstly
introduced SVM in 1995, with the main idea to ensure the network's high
generalization ability by mapping inputs non-linearly to high-dimensional
feature spaces, where linear decision surfaces were constructed with special
properties~\cite{CortesSupportvectornetworks1995}. The original idea of support
vector network was implied for the situation that training data was separable by
a hyperplane without error. Later, Cortes and Vapnik introduced the notion of
soft-margins such that a minimal subset of error in the training data is permit
table, allowing the remaining part of the training data to be separated by
constructing an optimal separating
hyperplane~\cite{CortesSupportvectornetworks1995}. Some advantages of the SVM
are the generalization of binary and regression forms and notation
simplification~\cite{BeggComputationalIntelligenceBiomedical2007}. SVM uses several
kernels such as the polynomial kernel, linear kernel, and the gaussian radial
basis function (RBF)
kernel~\cite{ScholkopfKernelMethodsComputational,KimFinancialtimeseries2003}.

\textbf{Decision Trees:} Decision Trees (DTs) are classifiers represented by a flowchart-like tree
structure introduced by J. R. Quinlan in
1986~\cite{QuinlanInductiondecisiontrees1986}. DTs do neither make any
statistical assumption concerning the inputs nor involve scaling of the data,
dissimilar to SVM and neural networks. DT models are constructed in terms of a tree
structure in which the dataset is broken down into smaller subsets at each
branch. Ultimately, the model results in a tree with decision nodes (branches)
and leaf nodes. DTs have been used for classification  in a variety of domains for pattern
recognition with its human-reasoning nature~\cite{AutomaticDesignDecisionTreea}.
Breiman has introduced the Classification and Regression Tree (CART) algorithm,
which allows continuous values to the model that can be used for regression
models~\cite{BreimanClassificationRegressionTrees1984}. The advantages of the decision
trees are self-explanatory logic flow, richness in representing discrete-value
classifier, and ability for handling data sets with error and missing data, while
the disadvantages are a shortage in classifier interaction and over-sensitivity
to irrelevant data and noise~\cite{RokachTopdowninductiondecision2005}.

\textbf{Artificial Neural Networks (ANNs):} ANNs are neural-network classifiers which simulate the function of the human
brain. They are a commonly used machine learning method. The network mainly
consists of three primary layers: the first layer represents input neurons; the
last layer represents output neurons; a series of weighted middle layers which
can minimize the error between actual output and forecasted
output~\cite{Wangcomparisonperformanceseveral2009}. It is difficult to extract
rules that ANNs set to interpret the model in the network since it is not to
analyze weights and bias terms in the network connections.

When we talk about ANNs we generally mean \emph{shallow networks} (less than 5
layers), in contrast to convolutional neural networks like VGG-19 that
are \emph{deep networks}.

%
%

\section{Image Dataset}\label{sec:dataset}
\begin{figure*}[htbp]
  \centering
  \vspace{0.1in}
  \includegraphics[width=0.95\textwidth]{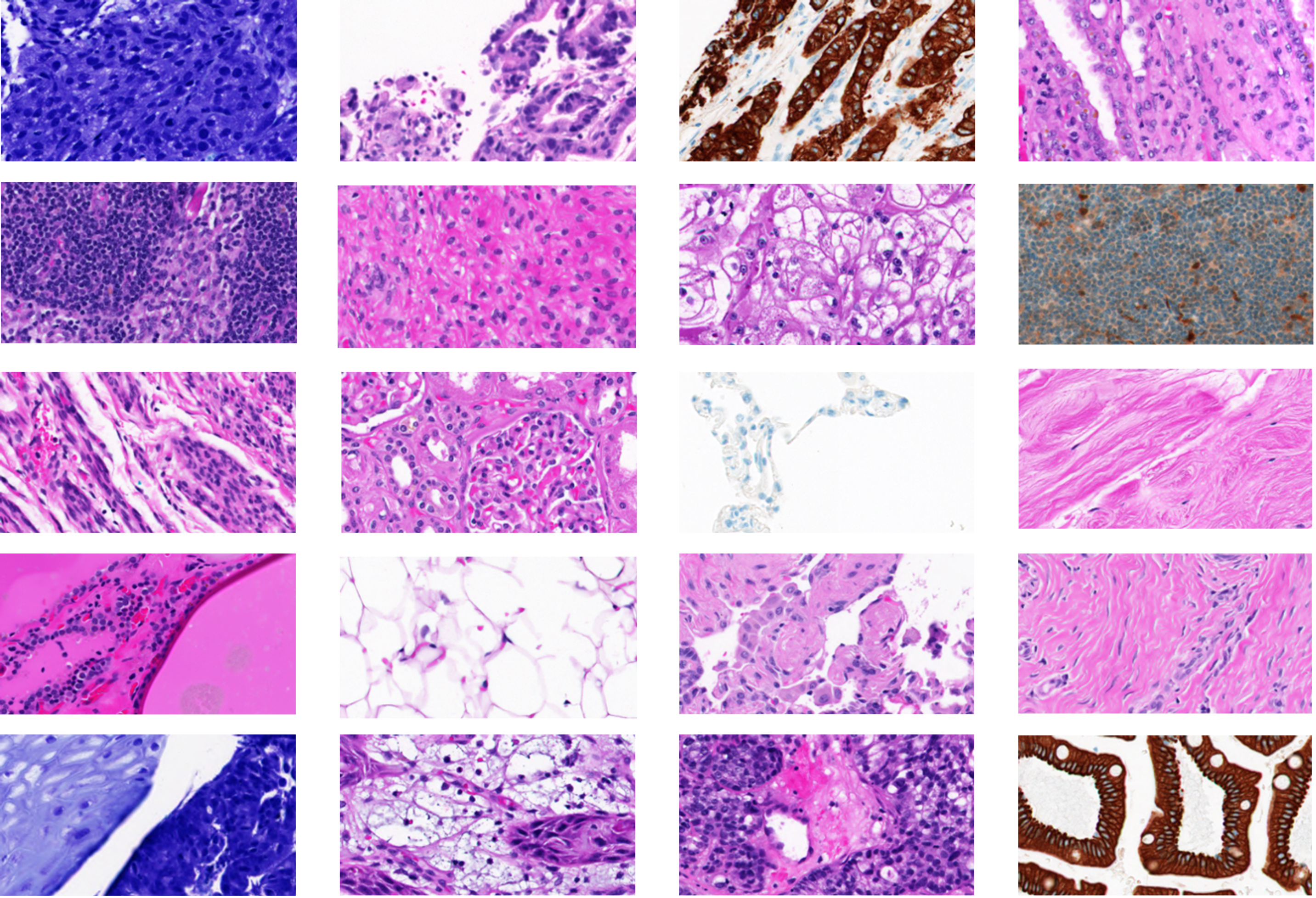}
  \caption{20 sample images, one from each class, showing the 20 classes of KIMIA Path 960 image dataset\label{fig:database}}
\end{figure*}

Histopathology images are used as the dataset in the this study obtained from
the KIMIA Lab\footnote{Source of dataset:
http://kimia.uwaterloo.ca/}, which contains 960
histopathology images that are collected from 400 whole slide images (WSIs) of
connective tissue, epithelial, and muscle in a colored TIF
format~\cite{KumarComparativeStudyCNN2017}. The dataset has 960 images, which
are obtained from 20 selected scans that visually represent different
texture/pattern types which are purely based on visual clues. These scans are of
the same size from 48 selected regions of interest from WSIs. The images are
down-sampled to $308 \times 168$. Fig.~\ref{fig:database} shows 20 sample
images, an image from each class, of the dataset, to illustrate the complexity
of the dataset as some of the classes have similar textures while others don't.

\subsection{Methodology}
For classifying the dataset, three models are constructed based on the features
extractor algorithms that are used. The models use LBP and HOG features
extractors, respectively.

The feature vectors obtained by each LBP, HOG or VGG19 features extracting
models are considered as inputs for the classifying methods, the feature
vectors are given independently to SVM, DTs, and ANNs. The accuracies are obtained
through the $k$-fold cross-validation method where $k = 3$.

Python programming language environment is used for conducting the experiments
with its supporting libraries provided by Anaconda distributions. Mahotas
library~\cite{CoelhoMahotasOpensource2013} is used for LBP features extractor.
Scikit-Image~\cite{vanderWaltscikitimageimageprocessing} library is used for HOG feature
extraction and SVM, DT, and ANN classification algorithms.

\subsection{Feature Extraction with HOG}
Image features extraction using HOG is applied in the first model. $18 \times
18$ cell size and $1 \times 1$ block size are used for computing the features.
The output of the HOG feature extraction is a histogram with $1224$ dimensions
(bins) for each image. The histograms can be used directly for training the
classifiers. A 3-fold cross-validation is performed to obtain the average for
SVM, DT, and ANN classification algorithms.

\subsection{Feature Extraction with LBP}
The second feature extraction model uses the LBP algorithm. The feature vectors
obtained through the application of LBP are captured in a histograms. The radius
parameter used is set to 4 (pixels) and the number of points to consider are set
to 14. The resulting histograms will be of 1182 dimensions (bins) for each
image. Therefore, these feature vectors can be used directly for training the
classifiers. A 3-fold cross-validation method have been used (with same random
initialization) in this model to obtain the accuracies for SVM, DT, and ANN.

\subsection{Deep Feature Extraction with VGG19}
Third feature extraction model uses the VGG19 deep network. Features for the
given pathology images are extracted from four different layers within
VGG19 network which contains 19 layers in total divided in group of 5 blocks.
These four layers are: \textit{fc1}, \textit{block5\_pool},
\textit{block4\_pool}, \textit{block3\_pool}. The reason for choosing multiple
locations is because the VGG19 network used for our experiments is pre-trained
on natural images using the ImageNet
dataset~\cite{KrizhevskyImagenetclassificationdeep2012}. The natural images
provide very different variability compared to the pathology images. However,
due to layered construction of VGG19, each layer is responsible for extracting
different type of features from an image, hence the deeper we go into the
network, the extracted features become more suitable for natural images and lose
their generality for other type of images such as histopathology images. We also
find this behaviour to be true, which will be discussed in
Section~\ref{sec:results}. 

We extracted 4 feature sets from the VGG-19 for Kimia
Path960 dataset, and without loosing generality, we performed same steps as
before in order to calculate accuracies for each of the classification models.
One may argue that, using ANNs on features extracted using VGG19 could be
considered as fine-tuning, however, for all our experiments, we are treating
classification methods separate from feature extraction models. Once each of our
feature extraction model provides its corresponding features, we apply same
classification methods on these features to calculate their discriminating
power.

\section{Experiments and Results}\label{sec:results}
\subsection{LBP Results}
The model that uses LBP features obtained the highest results in our
experiments. The accuracy is reported at 90.52\% using SVM which has a gamma
value of $0.0000015$ and penalty parameter of the error term $C = 2.5$ while
using the RBF kernel. The ANN classifier for LBP features consists of 300
neurons at the first and second layers and has a learning rate of $0.0005$. In
contrast, the DT classifier achieved an accuracy of $66.35\%$. The second row in
Table~\ref{tab:result} summarizes the accuracies obtained by the LBP features
extractor classification model using SVM, DT, and ANN methods (highest accuracy
is highlighted in bold).

\begin{figure}[htbp]
  \centering
  \includegraphics[width=1.1\columnwidth]{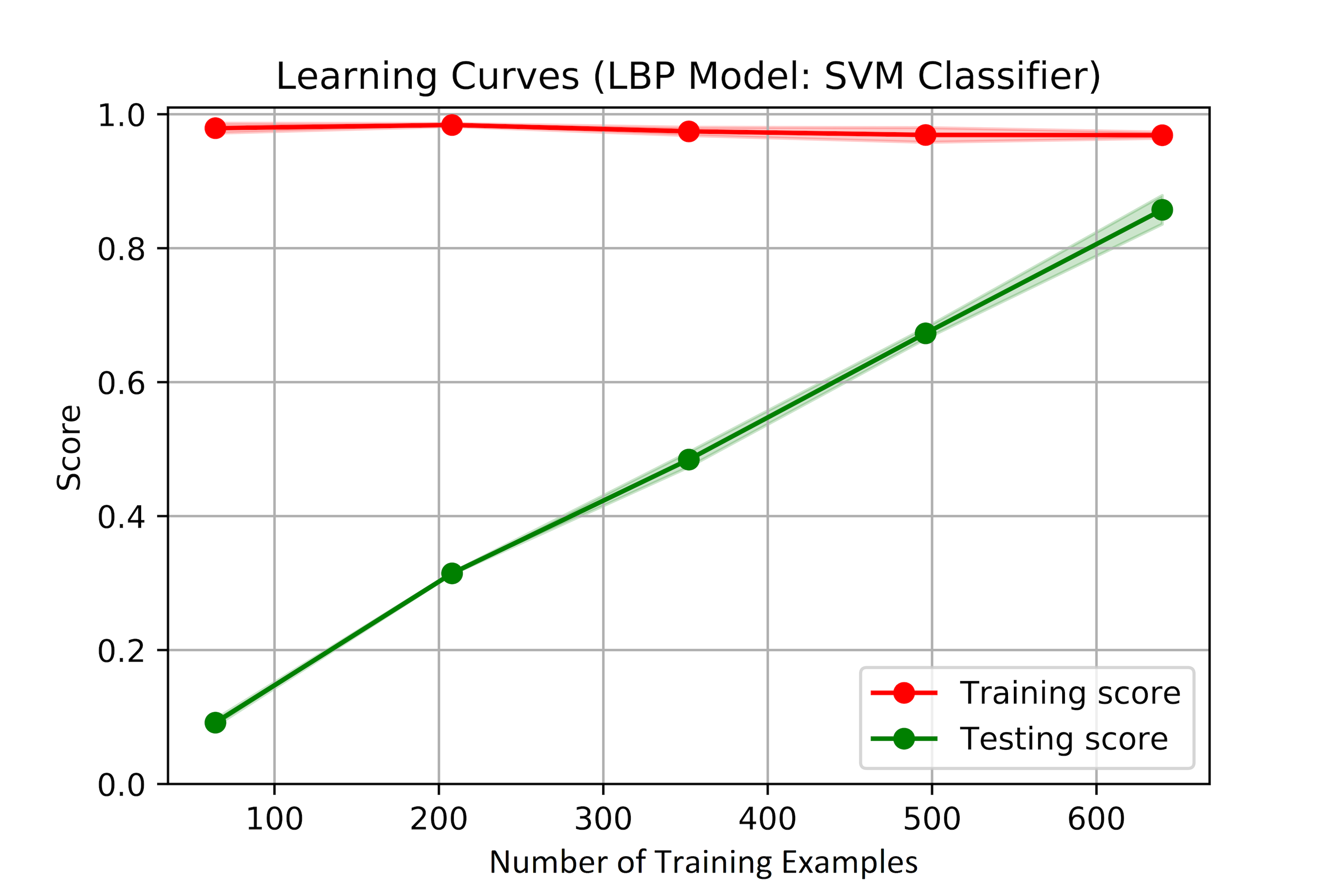}
  \caption{Plot of the SVM learning curves using LBP features. \label{fig:svm-lbp}}
\end{figure}

\begin{figure}[htbp]
  \centering
  \vspace{0.1in}
  \includegraphics[width=1.1\columnwidth]{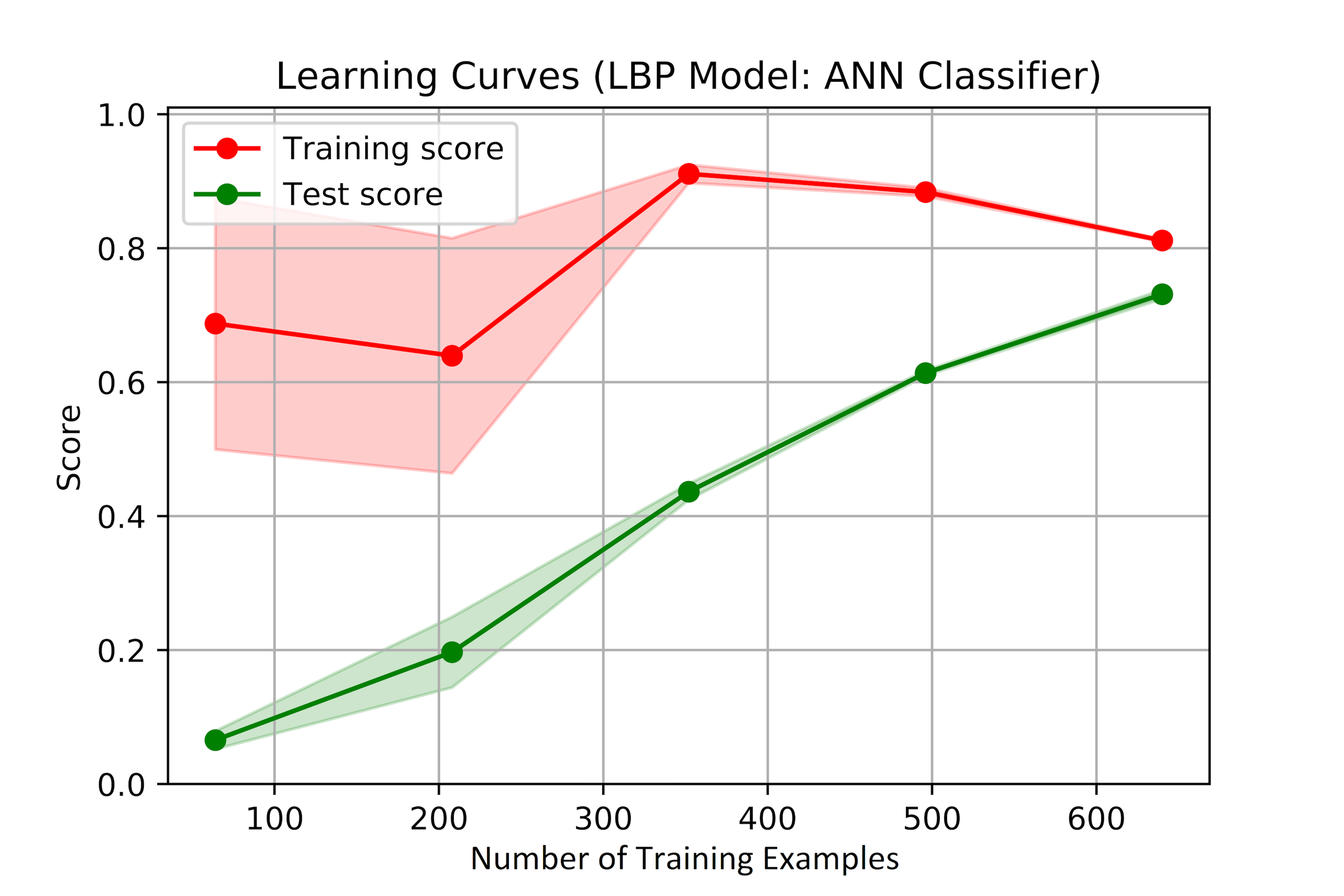}
  \caption{Plot of the ANN learning curves using LBP features.\label{fig:ann-classfier}}
\end{figure}

Fig.~\ref{fig:svm-lbp} and Fig.~\ref{fig:ann-classfier} plot the learning curves
of the SVM and ANN models that use LBP features. The lines in the
figures are the mean values of the scores and the highlighted area around the
lines is the range of its standard deviation, respectively. The red color is
used for training scores while the green color is used for the testing scores.

Fig.~\ref{fig:svm-lbp} shows that the training score was consistently high 
through the iterations. However, the testing scores were improving while the
training iterations were increasing. Fig.~\ref{fig:ann-classfier} illustrates
that the training score of the ANN classifier has reached its high-point at
around training iteration $500$. However, the testing score kept increasing till
it reaches roughly $660$ iterations.

\subsection{HOG Results}
The histograms obtained by HOG feature extraction consisted of $1224$ bins for
each image. The accuracies achieved by HOG features model are the lowest
compared to the other models. The SVM method resulted in $4.79\%$ accuracy. The
DT classifier computed $13.13\%$ accuracy and the ANN classifier achieved
$36.15\%$. All these classifiers had the same parameters as the classifier of
the LBP features classification model. Table~\ref{tab:result} summarizes the
accuracies obtained by the SVM, DT, and ANN classifiers with HOG features.

\subsection{Deep Features Results}
Feature vectors extracted from the later layers of VGG 19, for example
\textit{fc1} and \textit{block5\_pool}, do not perform well for the pathology
images. The highest accuracy on features obtained from these two layers is
achieved using DT classifier, $44.79$ and $47.18$, respectively. The reason DT
classifier dominates SVM or ANN is because of the high sparsity of the feature
vectors. Features from the \textit{block4\_pool} contribute to the highest
accuracy among deep features and are the second highest overall. However,
comparing the feature's dimensionality, its almost half the dimension of the
feature from LBP.

\begin{table*}[h]
\centering
\vspace{0.1in}
\input{result.table.tex}
\caption{The 3-fold cross-validation accuracies for LBP, HOG, VGG19
  features using SVM, DT, and ANN classifiers.\label{tab:result}}
\end{table*}

\section{Conclusion}\label{sec:conclusion}
The experiments conducted on KIMIA Path960 dataset using LPB, HOG and deep
network's features show that LPB feature extractor has outperformed the other
methods, especially while using SVM algorithm as a classifier. However, the
results obtained by HOG feature extractor were not satisfactory as the model has
under-performed by the parameters used in the present study. However, these
parameter choices were made to stay consistent in terms of feature's dimensions
across the multiple feature extractor models.

\begin{table*}[ht]
\centering
\input{result2.tex}
\caption{Show the best accuracies values, best classification methods and length of
  feature vectors for each feature extractor model. The best accuracy
  is achieved by LBP feature extractor using SVM classifier and second best is
  by VGG 19 (block4\_features) using SVM as well.\label{tab:final_comp}}
\end{table*}

Table~\ref{tab:final_comp} summarizes feature dimensions, classifier yielding
best accuracies and value of the best accuracies obtained by all three of the
feature extraction models.

The DTs have underperformed compared to SVM and ANN classifiers in most of the
cases expect when features were sparse, for example in case VGG 19
(block5\_pool) and VGG 19 (fc1) as in Table~\ref{tab:final_comp}. Meanwhile, SVM
algorithm has shown good results for overall expect for highly sparse features.
Therefore, one conclusion to be made is that selection between DT and SVM must
be made depending on the sparsity of feature vectors. The ANN algorithm had the
best accuracy using HOG features, which is the highest among all three
classifiers, otherwise ANN is classifier is generally under performing
with high standard deviation.

The accuracies obtained by HOG features extractor may be improved further by
changing the parameters of the algorithm. However, it required a lot of
resources to conduct these experiments. Furthermore, fine-tuning VGG 19 by
cutting it after block3\_pool may offer more improvements in accuracies on
pathology datasets as well.

Of course, that a handcrafted feature vector like LBP, with very simple
implementation, can beat deep features is a surprising result considering how
much efforts go into designing and training of a deep network. One may say that
deep nets need to be trained for the problem at hand with a large number of
training images. However, there are many situations where a large, balanced and
labelled dataset is not available.

\bibliographystyle{IEEEtran}
\bibliography{ref}
\end{document}

%% file: result.table.tex
\begin{tabular}{|c|c|ccc|}
\hline
\multicolumn{2}{|c|}{\multirow{2}{*}{\textbf{\begin{tabular}[c]{@{}c@{}}Feature Extraction \\ Models\end{tabular}}}} & \multicolumn{3}{c|}{\textbf{Classification Algorithms}}                                            \\ \cline{3-5} 
\multicolumn{2}{|c|}{}                                                                                               & \multicolumn{1}{c|}{\textbf{SVM}} & \multicolumn{1}{c|}{\textbf{DT}} & \textbf{ANN}                \\ \hline
\multirow{4}{*}{\textbf{LBP}}                                                                     & Fold 1           & $\mathbf{91.56\%}$                & $65.31\%$                        & $77.50\%$                   \\
                                                                                                  & Fold 2           & $\mathbf{88.75\%}$                & $66.25\%$                        & $76.25\%$                   \\
                                                                                                  & Fold 3           & $\mathbf{91.25\%}$                & $67.50\%$                        & $75.10\%$                   \\
                                                                                                  & All Folds        & $\mathbf{90.52 \pm 1.26\%}$       & $66.35 \pm 0.90\%$               & $75.10 \pm 2.56\%$          \\ \hline
\multirow{4}{*}{\textbf{HOG}}                                                                     & Fold 1           & $3.44\%$                          & $11.25\%$                         & $\mathbf{36.25\%}$          \\
                                                                                                  & Fold 2           & $7.81\%$                          & $13.44\%$                         & $\mathbf{35.94\%}$          \\
                                                                                                  & Fold 3           & $3.13\%$                          & $14.69\%$                         & $\mathbf{36.25\%}$          \\
                                                                                                  & All Folds        & $4.79 \pm 2.14\%$                 & $13.13 \pm 1.42\%$                & $\mathbf{36.15 \pm 0.15\%}$ \\ \hline
\multirow{4}{*}{\textbf{\begin{tabular}[c]{@{}c@{}}VGG19 \\ (fc1)\end{tabular}}}                  & Fold 1           & $9.37\%$                          & $\mathbf{45.31\%}$               & $5.62\%$                    \\
                                                                                                  & Fold 2           & $9.68\%$                          & $\mathbf{45.93\%}$               & $1.25\%$                    \\
                                                                                                  & Fold 3           & $9.06\%$                          & $\mathbf{43.12\%}$               & $4.37\%$                    \\
                                                                                                  & All Folds        & $9.37 \pm 0.25\%$                 & $\mathbf{44.79 \pm 1.2\%}$       & $7.50 \pm 1.83\%$           \\ \hline
\multirow{4}{*}{\textbf{\begin{tabular}[c]{@{}c@{}}VGG19 \\ (block5\_pool)\end{tabular}}}         & Fold 1           & $8.75\%$                          & $\mathbf{48.75\%}$               & $16.87\%$                   \\
                                                                                                  & Fold 2           & $5.93\%$                          & $\mathbf{47.50\%}$               & $18.12\%$                   \\
                                                                                                  & Fold 3           & $4.37\%$                          & $\mathbf{45.31\%}$               & $25.00\%$                   \\
                                                                                                  & All Folds        & $6.35 \pm 1.81\%$                 & $\mathbf{47.18 \pm 1.42\%}$      & $20.00 \pm 3.57\%$          \\ \hline
\multirow{4}{*}{\textbf{\begin{tabular}[c]{@{}c@{}}VGG19\\ (block4\_pool)\end{tabular}}}          & Fold 1           & $\mathbf{84.68\%}$                & $59.37\%$                        & $10.93\%$                   \\
                                                                                                  & Fold 2           & $\mathbf{80.31\%}$                & $58.43\%$                        & $17.50\%$                   \\
                                                                                                  & Fold 3           & $\mathbf{78.43\%}$                & $59.06\%$                        & $13.12\%$                   \\
                                                                                                  & All Folds        & $\mathbf{81.14 \pm 2.61\%}$       & $58.95 \pm 0.39\%$               & $13.85 \pm 2.73\%$          \\ \hline
\multirow{4}{*}{\textbf{\begin{tabular}[c]{@{}c@{}}VGG19\\ (block3\_pool)\end{tabular}}}          & Fold 1           & $\mathbf{67.81\%}$                & $62.81\%$                        & $2.81\%$                    \\
                                                                                                  & Fold 2           & $\mathbf{67.81\%}$                & $70.31\%$                        & $5.62\%$                    \\
                                                                                                  & Fold 3           & $66.87\%$                         & $\mathbf{72.18\%}$               & $6.25\%$                    \\
                                                                                                  & All Folds        & $67.49 \pm 0.44\%$                & $\mathbf{68.43 \pm 4.2 \%}$      & $4.90 \pm 1.49\%$           \\ \hline
\end{tabular}

%% file: result2.tex
\begin{tabular}{|cccc|}
\hline
\multicolumn{1}{|c|}{\textbf{\begin{tabular}[c]{@{}c@{}}Feature Extraction\\ Models\end{tabular}}} & \multicolumn{1}{c|}{\textbf{Feature Size}} & \multicolumn{1}{c|}{\textbf{\begin{tabular}[c]{@{}c@{}}Best Accuracy \\ Achieved\end{tabular}}} & \textbf{\begin{tabular}[c]{@{}c@{}}Best Classification\\ Method\end{tabular}} \\ \hline
\textbf{LBP}                                                                                       & 1182                                       & \textbf{90.52\%}                                                                                          & SVM                                                                           \\
\textbf{VGG 19 (block4\_pool)}                                                                     & 512                                        & 81.14\%                                                                                           & SVM                                                                           \\
\textbf{VGG 19 (block3\_pool)}                                                                     & 256                                        & 68.43\%                                                                                           & DT                                                                            \\
\textbf{VGG 19 (block5\_pool)}                                                                     & 512                                        & 47.18\%                                                                                           & DT                                                                            \\
\textbf{VGG 19 (fc1)}                                                                              & 4096                                       & 44.79\%                                                                                           & DT                                                                            \\
\textbf{HOG}                                                                                       & 1224                                      & 34.37\%                                                                                           & ANN                                                                           \\ \hline
\end{tabular}

%% file: paper.bbl
\ifdefined\DeclarePrefChars\DeclarePrefChars{'’-}\else\fi
\begin{thebibliography}{10}
\providecommand{\url}[1]{#1}
\csname url@samestyle\endcsname
\providecommand{\newblock}{\relax}
\providecommand{\bibinfo}[2]{#2}
\providecommand{\BIBentrySTDinterwordspacing}{\spaceskip=0pt\relax}
\providecommand{\BIBentryALTinterwordstretchfactor}{4}
\providecommand{\BIBentryALTinterwordspacing}{\spaceskip=\fontdimen2\font plus
\BIBentryALTinterwordstretchfactor\fontdimen3\font minus
  \fontdimen4\font\relax}
\providecommand{\BIBforeignlanguage}[2]{{%
\expandafter\ifx\csname l@#1\endcsname\relax
\typeout{** WARNING: IEEEtran.bst: No hyphenation pattern has been}%
\typeout{** loaded for the language `#1'. Using the pattern for}%
\typeout{** the default language instead.}%
\else
\language=\csname l@#1\endcsname
\fi
#2}}
\providecommand{\BIBdecl}{\relax}
\BIBdecl

\bibitem{deBruijneMachinelearningapproaches2016}
M.~de~Bruijne, ``Machine learning approaches in medical image analysis:
  {{From}} detection to diagnosis,'' vol.~33, pp. 94--97.

\bibitem{AIItsNature2016}
\emph{{{AI}}: {{Its Nature}} and {{Future}}}.\hskip 1em plus 0.5em minus
  0.4em\relax {Oxford University Press}.

\bibitem{GigerAnniversarypaperHistory2008}
M.~L. Giger, H.-P. Chan, and J.~Boone, ``Anniversary paper: {{History}} and
  status of {{CAD}} and quantitative image analysis: The role of {{Medical
  Physics}} and {{AAPM}},'' vol.~35, no.~12, pp. 5799--5820.

\bibitem{OliverFalsePositiveReduction2007}
A.~Oliver, X.~Lladó, J.~Freixenet, and J.~Martí, ``False {{Positive
  Reduction}} in {{Mammographic Mass Detection Using Local Binary Patterns}},''
  in \emph{Medical {{Image Computing}} and {{Computer}}-{{Assisted
  Intervention}} – {{MICCAI}} 2007}, ser. Lecture Notes in Computer
  Science.\hskip 1em plus 0.5em minus 0.4em\relax {Springer, Berlin,
  Heidelberg}, pp. 286--293.

\bibitem{CaicedoHistopathologyImageClassification2009}
J.~C. Caicedo, A.~Cruz, and F.~A. Gonzalez, ``Histopathology {{Image
  Classification Using Bag}} of {{Features}} and {{Kernel Functions}},'' in
  \emph{Artificial {{Intelligence}} in {{Medicine}}}, ser. Lecture Notes in
  Computer Science.\hskip 1em plus 0.5em minus 0.4em\relax {Springer, Berlin,
  Heidelberg}, pp. 126--135.

\bibitem{DoyleAutomatedgradingbreast2008}
S.~Doyle, S.~Agner, A.~Madabhushi, M.~Feldman, and J.~Tomaszewski, ``Automated
  grading of breast cancer histopathology using spectral clustering with
  textural and architectural image features,'' in \emph{2008 5th {{IEEE
  International Symposium}} on {{Biomedical Imaging}}: {{From Nano}} to
  {{Macro}}}, pp. 496--499.

\bibitem{KumarComparativeStudyCNN2017}
M.~D. Kumar, M.~Babaie, S.~Zhu, S.~Kalra, and H.~R. Tizhoosh, ``A {{Comparative
  Study}} of {{CNN}}, {{BoVW}} and {{LBP}} for {{Classification}} of
  {{Histopathological Images}}.''

\bibitem{OjalaPerformanceevaluationtexture1994}
T.~Ojala, M.~Pietikainen, and D.~Harwood, ``Performance evaluation of texture
  measures with classification based on {{Kullback}} discrimination of
  distributions,'' in \emph{Proceedings of 12th {{International Conference}} on
  {{Pattern Recognition}}}, vol.~1, pp. 582--585 vol.1.

\bibitem{HeTextureUnitTexture1990}
D.-c. He and L.~Wang, ``Texture {{Unit}}, {{Texture Spectrum}}, {{And Texture
  Analysis}},'' vol.~28, no.~4, pp. 509--512.

\bibitem{OjalaMultiresolutiongrayscalerotation2002}
T.~Ojala, M.~Pietikainen, and T.~Maenpaa, ``Multiresolution gray-scale and
  rotation invariant texture classification with local binary patterns,''
  vol.~24, no.~7, pp. 971--987.

\bibitem{JungLocalBinaryPatternBased2010a}
I.~Jung and I.~S. Oh, ``Local {{Binary Pattern}}-{{Based Features}} for {{Text
  Identification}} of {{Web Images}},'' in \emph{2010 20th {{International
  Conference}} on {{Pattern Recognition}}}, pp. 4320--4323.

\bibitem{WangHOGLBPhumandetector2009}
X.~Wang, T.~X. Han, and S.~Yan, ``An {{HOG}}-{{LBP}} human detector with
  partial occlusion handling,'' in \emph{2009 {{IEEE}} 12th {{International
  Conference}} on {{Computer Vision}}}, pp. 32--39.

\bibitem{BrahnamLocalBinaryPatterns2014}
``Local {{Binary Patterns}}: {{New Variants}} and {{Applications}}.''

\bibitem{DenizFacerecognitionusing2011}
O.~Déniz, G.~Bueno, J.~Salido, and F.~De~la Torre, ``Face recognition using
  {{Histograms}} of {{Oriented Gradients}},'' vol.~32, no.~12, pp. 1598--1603.

\bibitem{DalalHistogramsorientedgradients2005}
N.~Dalal and B.~Triggs, ``Histograms of oriented gradients for human
  detection,'' in \emph{2005 {{IEEE Computer Society Conference}} on {{Computer
  Vision}} and {{Pattern Recognition}} ({{CVPR}}'05)}, vol.~1, pp. 886--893.

\bibitem{SimonyanVeryDeepConvolutional2014}
K.~Simonyan and A.~Zisserman, ``Very {{Deep Convolutional Networks}} for
  {{Large}}-{{Scale Image Recognition}}.''

\bibitem{ErtosunProbabilisticvisualsearch2015}
M.~G. Ertosun and D.~L. Rubin, ``Probabilistic visual search for masses within
  mammography images using deep learning,'' in \emph{2015 {{IEEE International
  Conference}} on {{Bioinformatics}} and {{Biomedicine}} ({{BIBM}})}, pp.
  1310--1315.

\bibitem{KrizhevskyImagenetclassificationdeep2012}
A.~Krizhevsky, I.~Sutskever, and G.~E. Hinton, ``Imagenet classification with
  deep convolutional neural networks,'' in \emph{Advances in Neural Information
  Processing Systems}, pp. 1097--1105.

\bibitem{SzegedyGoingDeeperConvolutions2014}
C.~Szegedy, W.~Liu, Y.~Jia, P.~Sermanet, S.~Reed, D.~Anguelov, D.~Erhan,
  V.~Vanhoucke, and A.~Rabinovich, ``Going {{Deeper}} with {{Convolutions}}.''

\bibitem{WangDeepLearningIdentifying2016}
D.~Wang, A.~Khosla, R.~Gargeya, H.~Irshad, and A.~H. Beck, ``Deep {{Learning}}
  for {{Identifying Metastatic Breast Cancer}}.''

\bibitem{babaie2017local}
M.~Babaie, H.~Tizhoosh, A.~Khatami, and M.~Shiri, ``Local radon descriptors for
  image search,'' \emph{arXiv preprint arXiv:1710.04097}, 2017.

\bibitem{Tizhoosh2018}
H.~Tizhoosh and M.~Babaie, ``Representing medical images with encoded local
  projections,'' \emph{IEEE Transactions on Biomedical Engineering}, pp. 1--1,
  2018.

\bibitem{BeggComputationalIntelligenceBiomedical2007}
R.~Begg, D.~T. Lai, and M.~Palaniswami. Computational {{Intelligence}} in
  {{Biomedical Engineering}}.

\bibitem{CortesSupportvectornetworks1995}
C.~Cortes and V.~Vapnik, ``Support-vector networks,'' vol.~20, no.~3, pp.
  273--297.

\bibitem{ScholkopfKernelMethodsComputational}
B.~Schölkopf, K.~Tsuda, and J.-P. Vert. Kernel {{Methods}} in {{Computational
  Biology}}.

\bibitem{KimFinancialtimeseries2003}
K.-j. Kim, ``Financial time series forecasting using support vector machines,''
  vol.~55, no.~1, pp. 307--319.

\bibitem{QuinlanInductiondecisiontrees1986}
J.~R. Quinlan, ``Induction of decision trees,'' vol.~1, no.~1, pp. 81--106.

\bibitem{AutomaticDesignDecisionTreea}
\emph{Automatic {{Design}} of {{Decision}}-{{Tree Induction Algorithms}} |
  {{Rodrigo C}}. {{Barros}} | {{Springer}}}.

\bibitem{BreimanClassificationRegressionTrees1984}
L.~Breiman, J.~Friedman, C.~J. Stone, and R.~Olshen. Classification and
  {{Regression Trees}}.

\bibitem{RokachTopdowninductiondecision2005}
L.~Rokach and O.~Maimon, ``Top-down induction of decision trees classifiers - a
  survey,'' vol.~35, no.~4, pp. 476--487.

\bibitem{Wangcomparisonperformanceseveral2009}
W.-C. Wang, K.-W. Chau, C.-T. Cheng, and L.~Qiu, ``A comparison of performance
  of several artificial intelligence methods for forecasting monthly discharge
  time series,'' vol. 374, no.~3, pp. 294--306.

\bibitem{CoelhoMahotasOpensource2013}
L.~P. Coelho, ``Mahotas: {{Open}} source software for scriptable computer
  vision,'' vol.~1, no.~1, p.~e3.

\bibitem{vanderWaltscikitimageimageprocessing}
S.~van~der Walt, J.~L. Schönberger, and et. {al}. Scikit-image: Image
  processing in {{Python}} [{{PeerJ}}].

\end{thebibliography}
